\documentclass[11pt,a4paper]{article}
\usepackage{amsmath,amsfonts}

\def\hybrid{\topmargin -10pt    \oddsidemargin 0pt
        \headheight 0pt \headsep 0pt
       \textwidth 6.25in       
      \textheight 9.5in       
        \marginparwidth .875in
        \parskip 5pt plus 1pt   \jot = 1.5ex}
\hybrid
\begin{document}

\thispagestyle{empty}

\rightline{\small MIFPA-14-21}

\vskip 3cm
\noindent
{\LARGE \bf  Finite-temperature three-point function in 2D CFT}\\
\vskip .8cm
\begin{center}
\linethickness{.06cm}
\line(1,0){447}
\end{center}
\vskip .8cm
\noindent
{\large \bf Melanie Becker, Yaniel Cabrera, Ning Su}

\vskip 0.2cm

\hskip -0.1cm  {\em    George and Cynthia Mitchell Institute}
\vskip -0.15cm
{\em \hskip -.05cm for Fundamental Physics and Astronomy}
\vskip -0.15cm
{\em \hskip -.05cm Texas A \&M University, College Station}
\vskip -0.15cm
{\em \hskip -.05cm TX 77843--4242, USA}
\vskip -0.15cm
\vskip 0.5cm
{\tt \hskip -.5cm mbecker AT physics.tamu.edu, cabrera AT physics.tamu.edu, suning1985 AT gmail.com}

\vskip 1cm

\vskip0.6cm

\noindent
{\sc Abstract:}  We calculate the finite temperature three-point correlation function for primary fields in a 2D conformal field theory in momentum space. This result has applications to any strongly coupled field theory with a 2D CFT dual, as well as to Kerr/CFT.

\pagebreak

\newpage


With the discovery of AdS/CFT \cite{aharony99} and generalizations thereof like AdS/CMT \cite{iqbal11,sachdev11,herzog09,horowitz10}, Kerr/CFT \cite{guica09, bredberg11,compere12} and the applications of holographic duality to QCD \cite{casalderrey11}, the calculation of finite-temperature correlation functions in conformal field theory has become increasingly important.

 A lot of these holographic dualities contain black holes and thus describe dual field theories at finite temperature. For practical applications it is often convenient to work in momentum space. For example, even before the era of AdS/CFT it was noticed by Maldacena and Strominger \cite{maldacena97} that the semiclassical emission rates of scalar fields in a Kerr/Newmann black hole agree with the result obtained from a 2D effective conformal field theory in momentum space. The generalization to photons and fermions was later worked out by Gubser \cite{gubser97}.

In many cases the theory of interest is strongly coupled, so the conformal field theory calculation is the only reliable method to compute correlation functions. This is the point of view we take in this short note.

We are interested in finite temperature three-point functions for primary fields in 2D CFT in momentum space. First, we recall the calculation of the two-point function (see \cite{aharony99} and references therein). At zero temperature, the coordinate representation of this correlator for two primary fields $\phi_{h_i,\bar h_i}$ with conformal weights $(h_i,\bar h_i)$, $i =1 , 2$,  has the familiar form on the complex plane (after normalization)
\begin{equation}\label{two}
\langle \phi_{h_1,\bar h_1}(z,\bar z)\phi_{h_2,\bar h_2}(0,0)\rangle = \frac{\delta_{h_1, h_2}\delta_{\bar h_1,\bar h_2}}{z^{2 h_1}\bar z^{\bar2 h_1}}.
\end{equation}

We compactify this result on cylinders with radii $T_L$ and $T_R$ via the mapping
\begin{equation}
z=e^{2\pi T_L w}\ \ ,  \ \ \bar z=e^{2\pi T_R \bar w},
\end{equation}
where $T_L$, $T_R$, are the left and right moving temperatures. The imaginary part of $w= x + i\tau$ is the compactified time direction such that $\tau \sim \tau+ 1/T_L$. This mapping gives the finite-temperature two-point function in coordinate space

\begin{equation}\label{twoc}
\langle \phi_{h_1,\bar h_1}(\omega,\bar \omega)\phi_{h_2,\bar h_2}(0,0)\rangle =\left(\frac{\pi T_L}{\sinh \pi T_L w}\right)^{2h_1}\left(\frac{\pi T_R}{\sinh \pi T_R \bar w}\right)^{2\bar h_1},
\end{equation}
where we used $h_1=h_2$ and $\bar h_1=\bar h_2$; otherwise the correlator vanishes.

To compute the propagators we Wick rotate $\tau \rightarrow i t$ to obtain the light-cone coordinates $x^{\pm} = x\pm t$. We obtain the Lorentzian two-point function 
\begin{equation}\label{twocw}
\langle \phi_{h_1,\bar h_1}(x^+,x^-)\phi_{h_2,\bar h_2}(0,0)\rangle =\left(\frac{\pi T_L}{\sinh \pi T_L x^-}\right)^{2h_1}\left(\frac{\pi T_R}{\sinh \pi T_R  x^+ }\right)^{2\bar h_1}.
\end{equation}

Fourier transforming this expression gives the propagator in momentum space. We compute the retarded/advanced propagators, which reduce to the Euclidean Matsubara propagator when restricted to the discrete Matsubara frequencies $\omega = \pm 2\pi T k $, where $ k \in \mathbb{Z}$ \cite{son02}. To avoid singularities on the real axis the $\pm i\epsilon $ regularization prescription is implemented. This process leads to the advanced or  retarded propagators depending on the sign of $i\epsilon$.

For each sector we obtain \cite{gubser97, bredberg09}
\begin{equation}\label{twos}
G_\Delta ^{\pm}(\omega)=\displaystyle \int_{-\infty} ^\infty dx e^{-i \omega  x}\left(\frac{\pi  T}{\sinh (\pi  T x \pm i \epsilon)}\right)^{2\Delta }=(-1)^{\mp \Delta}\frac{(2\pi T)^{2\Delta-1}}{\Gamma(2 \Delta)}e^{\mp  \omega/2 T}\Gamma(\Delta + \frac{i \omega}{2\pi T})\Gamma(\Delta -\frac{i \omega}{2\pi T}).
\end{equation}

One can obtain Eq. (\ref{twos}) by means of the following Mellin-Barnes type integral \cite{bernd13}
\begin{equation}
\frac{1}{2 \pi  i}\int_{-i \infty }^{+i \infty } e^{\pm i\pi  s } \xi ^s  \Gamma (\Delta+s) \Gamma (\Delta-s) \, ds=\Gamma (2 \Delta) e^{\pm i \pi  \Delta } \xi ^\Delta(1-\xi \pm i \epsilon )^{-2 \Delta},
\end{equation}
where the integrand is defined on the principal branch of the complex logarithm.

After setting $\xi=e^{-2\pi T x}$, $ s = -i \omega /2\pi T$, and doing some manipulations, one obtains
\begin{equation}
\frac{1}{2\pi}\displaystyle \int _{-\infty} ^{\infty} {d\omega} e^{ i \omega x }(2\pi T)^{2\Delta - 1} \frac{1}{\Gamma(2\Delta)}e^{\pm \omega/2 T} \Gamma\left(\Delta-\frac{i\omega}{2 \pi T}\right)\Gamma\left(\Delta+\frac{i\omega}{2 \pi T}\right)=  e^{\pm i \pi  \Delta  } \left(\frac{\pi  T}{\sinh  (\pi  T x\pm i \epsilon )}\right)^{2 \Delta},
\end{equation}
which is the inverse of the Fourier transform of Eq. (\ref{twos}).

Next we calculate the momentum space finite temperature three-point function. Conformal symmetry fixes the form of the three-point function  up to an overall constant. At finite temperature, the holomorphic part is
\begin{equation}\label{corr}
\langle \phi_{h_1}(x^+_1)\phi_{h_2}(x^+_2)\phi_{h_3}(0)\rangle = \frac{(\pi T)^{l+m+n}}{\sinh^{m}(\pi T x^+_1)\sinh^{l}(\pi T x^+ _2)\sinh^{n}(\pi T x^+_{12})} .
\end{equation}
Here $m=h_3+h_1-h_2,  \ \ l=h_2+h_3-h_1, \ \ n=h_1+h_2-h_3$.

The holomorphic part of the momentum space three-point function follows by the Fourier transformation
\begin{equation}\label{fourier}
G_{m,n,l}(\omega_1,\omega_2) = \displaystyle \int_{-\infty} ^\infty \int_{-\infty} ^\infty dx_1 dx_2 \frac{(\pi T)^{l+m+n} \ e^{-i \omega_1 x_1 - i \omega_2 x_2}}{\sinh^m(\pi Tx_1)\sinh^l(\pi T x_2)\sinh^n(\pi T x_{12})}.
\end{equation}

Formally, the integral in Eq. (\ref{fourier}) is divergent so we work using the $i\epsilon$ prescription,
\begin{equation}\label{eps}
G_{m,n,l}(\omega_1,\omega_2) = \displaystyle \int_{-\infty} ^\infty \int_{-\infty} ^\infty dx_1 dx_2 \frac{(\pi T)^{l+m+n} \ e^{-i \omega_1 x_1 - i \omega_2 x_2}}{\sinh^m(\pi Tx_1+i\epsilon)\sinh^l(\pi T x_2+i\epsilon)\sinh^n(\pi T x_{12}-i\epsilon)}.
\end{equation}
The signs of the $i \epsilon$ terms are chosen such that the integral is convergent.

By introducing the variable $u=x_{12}$, we can disentangle the arguments of the hyperbolic sines in the integrand as
\begin{equation}\label{new}
\begin{split}
G_{m,n,l}(\omega_1,\omega_2) = \displaystyle \int_{-\infty} ^\infty \int_{-\infty} ^\infty \int_{-\infty} ^\infty & du dx_1 dx_2 \ \delta(u-x_1+x_2)\\
& \frac{(\pi T)^{l+m+n} \ e^{-i \omega_1 x_1 - i \omega_2 x_2}}{\sinh^m(\pi Tx_1+i\epsilon)\sinh^l(\pi T x_2+i\epsilon)\sinh^n(\pi T u-i\epsilon)}.
\end{split}
\end{equation}
Using the resolution of the identity,
\begin{equation}
\delta(x) = \displaystyle \frac{1}{2\pi} \int_{-\infty} ^\infty d\omega e^{-i \omega x},
\end{equation}
one can rewrite Eq. (\ref{new}) as an integral of a product of Fourier-transformed two-point correlators
\begin{equation}\label{product}
G_{m,n,l}(\omega_1,\omega_2) = \displaystyle \int_{-\infty} ^\infty \frac{d\omega}{2\pi} \ G_{m/2}^{+}(\omega_1-\omega) G_{n/2}^{-}(\omega) G_{l/2}^{+}(\omega_2+\omega),
\end{equation}
where, $G_\Delta ^{\pm}(\omega)$ is the Fourier transform of the two-point correlator with conformal weight $\Delta$ as given in Eq. (\ref{twos}).
It is easy to see that the above expression agrees with the one for the extremal three-point correlation function computed in \cite{becker11}, when either $m=0, n=0$, or $l=0$. It also reduces to the advanced or retarded propagator by setting one of the conformal weights to zero and the remaining two equal to each other. For example, if one chooses $h_3 = 0$ and $h_1=h_2$ then
\begin{equation}
G_{0,2h_1,0}(\omega_1,\omega_2) = \delta(\omega_1+\omega_2)G^{-}_{h_1}(\omega_1).
\end{equation}

 We proceed by substituting the expression for $G^{\pm} _\Delta$ into Eq. (\ref{product}),

\begin{equation}\label{mid}
\begin{split}
G_{m,n,l}&(\omega_1,\omega_2) =(-1)^{(-m+n-l)/2} \frac{(2\pi T)^{m+n+l-3}}{\Gamma( m)\Gamma( n)\Gamma( l)} e^{(\omega_1+\omega_2)/2T}
 \displaystyle \int_{-\infty}^\infty \frac{d\omega}{2\pi}\ \left( e^{-\omega/2T}\Gamma(\frac{m}{2}+i \frac{\omega_1}{2\pi T}-i\frac{\omega}{2\pi T})\right. \\
&\left. \ \ \ \ \ \Gamma(\frac{m}{2}- i \frac{\omega_1}{2\pi T}+ i\frac{\omega}{2\pi T})\Gamma(\frac{n}{2}+i \frac{\omega}{2\pi T})\Gamma(\frac{n}{2}-i \frac{\omega}{2\pi T}) \Gamma(\frac{l}{2}+i\frac{\omega_2}{2\pi T}+i\frac{\omega}{2\pi T})\Gamma(\frac{l}{2}-i\frac{\omega_2}{2\pi T}-i\frac{\omega}{2\pi T}) \right).
\end{split}
\end{equation}

With the simple change of variables $\omega=i 2\pi T s$ we recognize the above integral as the defining integral of the Meijer-G function on the first path \cite{erderlyi, gradshteyn}
\begin{equation}\label{meijer2}
\begin{split}
G_{m,n,l}(\omega_1,\omega_2) =(-1)^{(m-n+l)/2}& \frac{( 2\pi T)^{m+n+l-2}}{\Gamma( m)\Gamma( n)\Gamma( l)} e^{-(\omega_1+\omega_2)/2T}\\ &G_{3,3}^{3,3}\left(
\begin{array}{c}
 1-\frac{m}{2}-i\frac{\omega_1}{2\pi T},1-\frac{n}{2},1-\frac{l}{2}+i \frac{\omega_2}{2\pi T} \\
 \frac{m}{2}-i\frac{\omega_1}{2\pi T},\frac{n}{2},\frac{l}{2}+i \frac{\omega_2}{2\pi T} \\
\end{array}
,e^{-i \pi }\right).
\end{split}
\end{equation}
One easily checks that the convergence conditions for this function are satisfied because $p+q<2(r+s)$, $| \arg z | < ( r + s -(p+q)/2)\pi$ for $G_{r,s}^{p,q}$ with argument $z$, and the poles of the Gamma functions are on the right locations \cite{gradshteyn}.

For general values of $m, n$ and $l$ this function has to be evaluated numerically. For integer values of these parameters the function can be evaluated analytically. For example, with $m =1, n =3$ and $l = 2$ we get from Eq. (\ref{meijer2})
\begin{equation}\label{example}
\begin{split}
G_{1,3,2}(\omega_1,\omega_2) = &\frac{\pi ^2}{6 \left(e^{-\frac{\omega _2}{T}}+1\right) \left(1-e^{-\frac{\omega _1}{T}}\right)} \left(\pi ^2 T^2+\omega _2^2\right) \left(9 \pi ^2 T^2+\omega _2^2\right)+\\&\frac{\pi ^2}{6 \left(e^{\frac{-\omega _1-\omega_2}{T}}+1\right) \left(1-e^{\frac{\omega _1}{T}}\right)} \left(\pi ^2 T^2+(\omega _1+\omega_2)^2\right) \left(9 \pi ^2 T^2+3\omega_1 ^2 +\omega_2 ^2 -2\omega_1\omega_2\right).
\end{split}
\end{equation}

We see that the finite-temperature three-point function has a similar frequency dependence as the greybody factors found in the absorption cross-section in \cite{maldacena96}. It is also easy to check that the zero temperature limit of Eq. (\ref{example}) precisely gives the Fourier transform of the correlation function on the plane, 
\begin{equation}
\lim_{T\rightarrow 0} G_{1,3,2}(\omega_1,\omega_2) = \int_{-\infty}^\infty dx_1 dx_2 e^{-i\omega_1 x_1-i\omega_2 x_2 }\frac{1}{(x_1+i\epsilon)^1}\frac{1}{(x_2+i\epsilon)^2}\frac{1}{(x_1-x_2 -i\epsilon)^3}.
\end{equation}

The result in Eq. (\ref{meijer2}) holds for any (even strongly coupled) field theory with a 2D CFT holographic dual. It would be interesting to see if this result agrees with the bulk calculation of three-point functions in the near-NHEK geometry, as proposed by the Kerr/CFT correspondence \cite{guica09, bredberg11}, beyond the extremal case considered in \cite{becker11}. Work in this direction is in progress.

\section*{Acknowledgements}

We thank Katrin Becker, Tom Hartman, and Andrew Strominger for useful discussions. This work was supported by the NSF under the grants PHY-0505757, DMS-1159404 and  DGE-1252521 and by Texas A\&M University.

\bibliographystyle{utphys}
\bibliography{kerrcftscalar3pt}

\end{document}